\documentclass[article,twocolumn, showpacs,
]{revtex4}
\newcommand{\be}{\begin{equation}}
\newcommand{\ee}{\end{equation}}
\newcommand{\bea}{\begin{eqnarray}}
\newcommand{\eea}{\end{eqnarray}}
\usepackage{graphicx}
\usepackage{epsfig}
\usepackage{bm}

\usepackage{amssymb}

\begin{document}
\title{Graphene Conductivity near the Charge Neutral Point}
\author{L. Moriconi}
\author{D. Niemeyer}
\affiliation{Instituto de F\'\i sica, Universidade Federal do Rio de Janeiro, \\
C.P. 68528, 21945-970, Rio de Janeiro, RJ, Brazil}
\begin{abstract}
Disordered Fermi-Dirac distributions are used to model, within a straightforward and essentially
phenomenological Boltzmann equation approach, the electron/hole transport across graphene puddles. 
We establish, with striking experimental support, a functional relationship between the graphene 
minimum conductivity, the mobility in the Boltzmann regime, and the steepness of 
the conductivity parabolic profile usually observed through gate-voltage scanning around the charge 
neutral point.
\end{abstract}
\maketitle

\section{Introduction}

A great deal of effort has been currently devoted to the study of the transport properties of graphene, with particular focus on the so-called minimum conductivity problem, i.e., the fact that monolayer graphene's conductivity drops to a finite minimum value when the Fermi level is set at the charge-neutral point, where the density of states should hypothetically vanish (for non-disordered and non-interacting electrons) \cite{geim_novo, castro_etal, peres, mucci-caio, DasSarma_etal}. A careful attempt to use standard linear response theory in this context is, however, plagued by ambiguities related to the alternative prescriptions that can be used in the computation of quantum matrix elements \cite{ryu_etal}. Notwithstading such puzzling theoretical issues, it is generally accepted, on clear experimental grounds, that the minimum conductivity, for both mono and bilayer graphene, is a non-universal quantity (frequently observed to be around $4 e^2/h$), being essentially dependent on the amount and the nature of electronic disorder.

It is a spread view that a solution of the minimum conductivity problem has to do more with finding appropriate models of disorder than with improving electronic transport theory. In this paper, we rely on the ``puddle picture"  of disorder, which is known to yield a consistent explanation of the finite minimum conductivity phenomenon for a large class of graphene systems \cite{DasSarma_etal, hwang_etal, adam_etal, martin_etal, zhang_etal,DasSarma2_etal, deshpande_etal}. The essential idea of the puddle picture is that the disorder produced by the sample and substract impurities is associated to local and smooth shifts of the electron energy spectrum. Due to the uniqueness of the Fermi level, the electric charge is distributed in disordered graphene in the form of negatively (electrons) or positively (holes) charged puddles. Furthermore, it is believed, in connection with the celebrated Klein tunneling effect \cite{kats_etal}, that electric charge can be transported across the puddles with strongly supressed backscaterring. At the charge neutral point, graphene is then depicted as a mixture of electrons and holes, a fact that was actually pointed out in the very first graphene transport experiments \cite{novo_etal}.

We note, as an important remark, that the graphene conductivity has a non-vanishing minimum value even in clean suspended samples, where charged puddles should not exist due to the absence of a substract. Thus, the puddle picture cannot be evoked in these cases, where it is likely that more general quantum phenomena play a relevant role in the transport process \cite{kats,drago,trushin_etal}. 

It is clear that a comprehensive test of the puddle picture should address, with reasonable accuracy, the characterization of disorder in the graphene samples under investigation. Having in mind that this is a modeling task of difficult validation, it would be of great interest to establish general results that would not rely on most of the disorder details. This is our aim in this work, to be pursued within the framework of the Boltzmann approach, however through an alternative implementation of the puddle picture of disorder.

This paper is organized as follows. In Sec. II, we rephrase the puddle model of disordered graphene as a network of coupled fermion systems with randomly shifted energy spectra. This system is studied through the usual Boltzmann transport theory in Sec. III, where our main result -- a statement on the behavior of conductivity near the charge neutral point -- is derived and clearly confirmed from available experimental data. In Sec. IV we discuss, with the help of standard phenomenological ideas, why the application of the Boltzmann formalism is indeed meaningful in the analysis of the conductivity profile around the charge neutral point, a fact that could seem paradoxical at first sight, since near the Dirac point the Fermi wavelength is large enough to break the semiclassical transport regime. In Sec. V, we summarize our findings and point out directions of further research.

\section{Statistical Description of Graphene Puddles}

Consider a system of gapless fermions with one-particle spectrum $\epsilon_k  = a k^\alpha$ ($\alpha=1$ and $\alpha=2$ correspond to ideal monolayer and bilayer graphene systems, respectively). As sketched in Fig. 1, we depict graphene electron-hole puddles as open ideal, two-fluid subsystems sharing chemical potential $\mu$ for electrons and $-\mu$ for holes, in a large charge transport network \cite{cheianov_etal}. The role of disorder is encoded in the ``energy broadening function" $\rho(\xi)$, which is just the probability density function of finding a subsystem with shifted electron or hole energy spectrum $\epsilon_k - \xi$ in a given puddle. The energy broadening width $\delta \epsilon_0$ is 
defined from
\be
(\delta \epsilon_0)^2  = \int_{- \infty}^\infty d \xi \xi^2 \rho (\xi)  \ , \ \label{en-broaden}
\ee
and is assumed to yield a complete parametrization of $\rho(\xi)$. In other words,
we have
\be
\rho(\xi) =  \frac{1}{\delta \epsilon_0} g(\xi / \delta \epsilon_0) \ , \ \label{rho}
\ee
where $g( \cdot )$ is a dimensionless universal probability density function (an educated 
guess is to take it as a zero-mean gaussian distribution with unit standard deviation, but
its exact form is not important in our discussion).

\begin{figure}[tbph]
\vspace{-5.5cm}
\includegraphics[width=8.0cm, height=11.36cm]{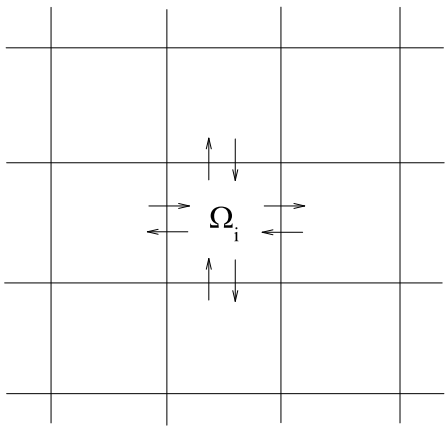}
\caption{Schematic view of disordered graphene as a network of 
open smaller ideal subsystems $\Omega_i$ (electron-hole puddles), which have 
one-particle energy spectra $\epsilon_k - \xi_i$, where the $\xi_i$'s are independent 
and identically distributed random variables. Arrows indicate the flow of particles and 
holes through the boundaries of $\Omega_i$.}
\label{}
\end{figure}

In accordance with the statistical picture of electron-hole puddles advanced here, we put 
forward the zero-temperature equilibrium electron/hole occupation numbers at wavenumber 
$\vec k$ as
\be
f_0^\pm(\vec k) = \int_{- \infty}^\infty d \xi \rho(\xi) \Theta(\xi \pm \mu - \epsilon_k) \label{f} 
\ , \
\ee
where $\Theta( \cdot )$ is the Heaviside function, and the above positive and negative indices refer 
to electrons and holes, respectively. We have tacitly assumed, in Eq. (\ref{f}), that fermions are 
described by an electron-hole symmetric hamiltonian \cite{comment}. 

Observe that the total electric charge
\be
Q = e \left ( \frac{L}{2 \pi} \right )^2 \int d^2 \vec k [f_0^+(\vec k) - f_0^-(\vec k)] \ , \
\ee
vanishes, from Eq. (\ref{f}), at $\mu=0$, the so-called charge neutral point.
On the other hand, the carrier density (i.e., the mean number of electrons and holes per unit area) 
at the charge neutral point, $n_0$, is finite, and is readily evaluated as
\bea
&&n_0=n_+(0)+n_-(0) = \nonumber \\
&&= \frac{2}{(2 \pi)^2} \int_{-\infty}^\infty d \xi \rho(\xi) \int d^2 \vec k \Theta(\xi - a k^\alpha) \nonumber \\
&&=  \frac{1}{2 \pi} \left (\frac{\delta \epsilon_0}{a} \right )^{\frac{2}{\alpha}} \int_0^\infty d \xi g(\xi) 
\xi^{\frac{2}{\alpha}}  \ . \ \label{n0}
\eea

\section{Conductivity Profile Around the Charge Neutral Point}

Under the action of a small external electric field $\vec E$, time-dependent occupation numbers  
can be obtained in principle as solutions of the usual Boltzmann equations in the relaxation time 
approximation, viz.,
\be
[\frac{\partial}{\partial t} \pm \frac{e \vec E}{\hbar} \cdot \vec \nabla_k ]f^\pm(\vec k,t) = - \frac{1}{\tau_k} 
[f^\pm(\vec k,t) - f^\pm_0(\vec k)] \ , \ \nonumber \\
\label{b-eqs}
\ee
where $\tau_k$ is the scattering time at wavenumber $k$. The stationary solutions of (\ref{b-eqs}) are, up to 
first order in the electric field,
\be
f^\pm(\vec k) = \left [ 1 \mp \frac{e \vec E \tau_k}{\hbar} \cdot \vec \nabla_k \right ] f^\pm_0(\vec k)  \ . \ 
\label{f_st}
\ee
Substituting (\ref{f_st}) in the expression of the electric current density,
\be
\vec j = \frac{\alpha a e}{\hbar} \left ( \frac{1}{2\pi} \right )^2 \int d^2 \vec k  [f^+(\vec k) - f^-(\vec k)]
k^{\alpha -1} \hat k \ , \ \label{int_v}
\ee
we find the conductivity
\be
\sigma =   \frac{\alpha \pi e^2}{h^2} T(\mu) \ , \ \label{sigma-t}
\ee
where
\be
T(\mu) \equiv \int_0^\infty d \xi [\rho(\xi+\mu)+\rho(\xi - \mu)] \xi \tau_{k_\xi} \ , \ \label{t-mu}
\ee
with $k_\xi = (\xi / a)^{1/\alpha}$. 

The scattering time at wavenumber $k$ is expected to have the general form $\tau_k = F(k) / n_{imp}$, 
where $n_{imp}$ is the impurity concentration and $F(k)$ is some scaling function of $k$ \cite{DasSarma_etal, adam_etal}. 
Rather than attempting to compute $F(k)$ on a first-principle basis, we determine how it should scale with $k$ from 
the observed behavior of conductivity. More concretely, we assume that far from the charge neutral 
point, that is, for $|\mu|/ \delta \epsilon_0 \gg 1$, the conductivity becomes a linear function of the 
carrier density $n$. In this region, (i) the chemical potential can be identified to the Fermi energy of an 
ideal gas, that is, $|\mu| = a k^\alpha$, and (ii) Eqs. (\ref{sigma-t}) and (\ref{t-mu}) yield 
$\sigma = \alpha \pi e^2 \mu \tau_{k_\mu} / 2 h^2$. Since the carrier density is $n \propto k^2$, the 
conductivity will depend linearly on $n$ only if $F(k) \propto k^{2 - \alpha}$. Therefore, we may
conventionally write the scattering time as
\be
\tau_k = \frac{c \hbar k^{2-\alpha}}{ \alpha a n_{imp}} \ , \  \label{tau-k}
\ee 
where $c$ is a dimensionless prefactor. It is worth of mentioning that Eq. (\ref{tau-k}), which is taken 
to hold also around the charge neutral point, can be in fact derived for monolayer graphene ($\alpha=1$) 
in the case of Coulomb impurity potentials \cite{DasSarma_etal, adam_etal}. 

In order to investigate the conductivity behavior close to the charge neutral point, let us expand (\ref{t-mu}) 
around $\mu = 0$. We have, up to second order in $\mu$,
\be
T(\mu) = 2\int_0^\infty d \xi \xi \tau_{k_\xi} [\rho(\xi) + \frac{\mu^2}{2}  \rho''(\xi) ] \ . \ \label{t-mu2}
\ee
Substituting (\ref{tau-k}) in (\ref{t-mu2}) and taking (\ref{rho}) 
into account, we obtain
\bea
&&T(\mu) = \frac{2c \hbar }{\alpha n_{imp}} \left ( \frac{\delta \epsilon_0}{a} \right  )^\frac{2}{\alpha} \times \nonumber \\
&&\int_0^\infty d \xi \xi^{\frac{2}{\alpha}} \left [g(\xi) + \frac{1}{2} \left ( \frac{\mu}{\delta \epsilon_0} \right )^2 g''(\xi) \right ] \ , \
\eea
so that
\be
\sigma = \sigma_0 + \frac{\sigma_1}{2} \left ( \frac{\mu}{ \delta \epsilon_0} \right ) ^2 + {\cal{O}}(\mu^4) \ , \ \label{sigma_mu}
\ee
where
\be
\sigma_0 = \frac{e^2}{h} \frac{c }{n_{imp}} \left ( \frac{\delta \epsilon_0}{a} \right  )^\frac{2}{\alpha}
\int_0^\infty d \xi  g(\xi) \xi^{\frac{2}{\alpha}}  \label{sigma_0}
\ee
is the minimum conductivity, and
\be
\sigma_1 = \sigma_0 \frac{\int_0^\infty d \xi  g''(\xi) \xi^{\frac{2}{\alpha}}}{\int_0^\infty d \xi  g(\xi) \xi^{\frac{2}{\alpha}}} \ . \ \label{sigma_1}
\ee

It is usual to control the two-dimensional carrier charge density in transport experiments through the voltage bias provided by gate electrode devices \cite{geim_novo}. Variations of the gate voltage $V_g$ are proportional to charge density variations $c_g \Delta V_g$, where $c_g$ is the gate capacitance per unit area. Let us take, without loss of generality, $V_g=0$ at the charge neutral point. We can relate $\mu$ to $V_g$ from the expression of the total charge density $e[n_+(\mu)-n_-(\mu)]$. With the help of Eq. (\ref{f}), we get, up to first order in $\mu$,
\bea
&&n_+(\mu)-n_-(\mu) = \frac{1}{(2 \pi)^2} \int_{-\infty}^\infty d \xi \rho(\xi) \int d^2 \vec k  \nonumber \\
&&\times [\Theta(\xi + \mu - \epsilon_k)
- \Theta(\xi - \mu - \epsilon_k)] \nonumber \\
&&= \frac{1}{4 \pi} \int_0^\infty d \xi  [\rho(\xi -\mu)- \rho(\xi + \mu)] \left (\frac{\xi}{a} \right )^{\frac{2}{\alpha}} \nonumber \\
&&= - \frac{\mu}{4 \pi} \int_0^\infty d \xi \rho'(\xi) \left (\frac{\xi}{a} \right )^{\frac{2}{\alpha}}  + {\cal{O}}(\mu^2) \nonumber \\
&&= - \frac{\mu}{4 \pi a} \left (\frac{\delta \epsilon_0}{a} \right )^{\frac{2}{\alpha}-1}\int_0^\infty d \xi g'(\xi) \xi^{\frac{2}{\alpha}}+ {\cal{O}}(\mu^2) 
\ . \
\eea
The above result implies, thus, that for $|\mu| \ll \delta \epsilon_0$,
\be
V_g = - \frac{e \mu}{c_g 4 \pi a} \left (\frac{\delta \epsilon_0}{a} \right )^{\frac{2}{\alpha}-1}\int_0^\infty d \xi g'(\xi) \xi^{\frac{2}{\alpha}}
\ . \  \label{vg-mu}
\ee

\begin{figure}[tbph]
\includegraphics[width=9.68cm, height=6.55cm]{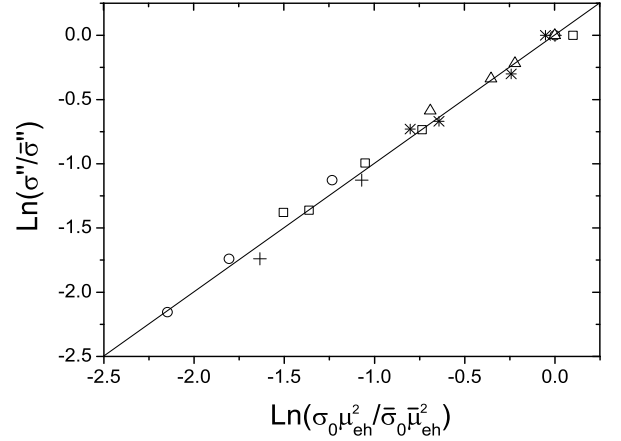}
\caption{The dimensionless steepness, $\sigma'' / \bar \sigma''$, of the graphene conductivity parabola near the charge neutral point 
is compared to the dimensionless combination of minimum conductivity and mobility, $\sigma_0 \mu_{eh}^2 / \bar \sigma_0 \bar \mu_{eh}^2$. 
Open circles and crosses refer to electron and hole mobilities, respectively \cite{chen_etal}. The other symbols are related to non-specified electron or hole mobilities. Data source: triangles \cite{chen2_etal}; squares \cite{mccreary_etal}; asterisks (bilayer graphene) \cite{xiao_etal}. The squares and asterisks are slightly shifted to the right and to the left, respectively, to help visualization of the linear interpolation, which has unit slope as predicted by (\ref{sigma-mu_eh}).}
\label{}
\end{figure}

We have at this point all the ingredients to establish a phenomenological theorem for the steepness of the conductivity 
parabolic profile around $V_g=0$, viz.,
\be
\left. \sigma'' \equiv \frac{\partial^2 \sigma}{\partial V_g^2} \right |_{V_g = 0} \propto \sigma_0 \mu_{eh}^2 \ , \ \label{sigma-mu_eh}
\ee
where  $\mu_{eh} = \sigma / en \propto 1 / n_{imp}$ is the carrier mobility measured far from the charge neutral point. 
Relying now upon the fact that at the charge neutral point free charge carriers come from donor or acceptor impurities, we 
have $n_0 \propto n_{imp}$ \cite{zhang_etal}, so that $\mu_{eh} \propto 1 / n_0$. Relation (\ref{sigma-mu_eh}), then, follows 
immediately from (\ref{n0}), (\ref{sigma_mu})-(\ref{sigma_1}), and (\ref{vg-mu}).

We have taken $\sigma''$, $\sigma_0$ and $\mu_{eh}$ from the experimental data reported in previous studies \cite{chen_etal, chen2_etal, mccreary_etal,xiao_etal}. For each group of measurements, carried out for various impurity concentrations, we denote by $\bar \sigma''$ and  $\bar \sigma_0$ the parabola steepness and the minimum conductivity associated to the largest selected mobility $\bar \mu_{eh}$. We work, then, with the dimensionless quantities $\sigma'' / \bar \sigma''$ and $\sigma_0 \mu_{eh}^2 / \bar \sigma_0 \bar \mu_{eh}^2$ to find, as it is shown in Fig. 2, compelling evidence for the validity of (\ref{sigma-mu_eh}).

It is interesting to check, additionally, if the statistical description of graphene transport that has led ultimately to (\ref{sigma-mu_eh}) is also able to provide a reasonable estimate of the minimum conductivity value. Note that with the help of Eqs. (\ref{n0}) and (\ref{sigma_0}), we find
\be
\sigma_0 = 2 \pi c  \frac{n_0}{n_{imp}}  \frac{e^2}{h}  \ . \ \label{sigma_est}
\ee
The dimensionless constant $c$ is, in the case of Coulomb impurities,  a function of the Wigner-Seitz radius $r_s$,
\be
c = \frac{1}{4 \pi G(r_s)}  \  , \
\ee
where \cite{DasSarma_etal,adam_etal}, 
\be
G(r_s) = r_s^2 \left \{ \frac{\pi}{2} - 4 \frac{d}{d r_s} [ r_s^2 g(2 r_s) ] \right \} \ , \
\ee
with
\be
g(x) = -1 + \frac{\pi}{2} x + (1-x^2) \frac{1}{\sqrt{x^2-1}} \arccos \frac{1}{x} \ . \ \label{g-func}
\ee
Considering SiO$_2$ as the prototypical substract, we take, as estimates for monolayer graphene, 
$r_s = 0.8$ \cite{DasSarma_etal, adam_etal} and $n_0 =  n_{imp}/10$ \cite{zhang_etal}. It follows, 
from Eqs. (\ref{sigma_est})-(\ref{g-func}) that these two numerical values conspire to give 
$\sigma_0 = 1.00834 \times e^2 / h$, which is incidentally very close to half of the conductance 
quantum. Taking into account the double valley and the spin 1/2 degrees of freedom of graphene 
electrons, its minimum conductivity is expected to be around $4 e^2 /h$, as in fact it has been 
reported in several transport experiments. To be fair, however, we stress that the measured 
$\sigma_0$'s are strongly sample-dependent \cite{geim_novo,chen2_etal}, varying typically in 
the range $(2-7)e^2/h$. As we clarify in the next section, our evaluation of the minimum 
conductivity has to be interpreted more as an order of magnitude estimation than as a procedure 
to obtain a numerically precise result.

\section{Critical Remarks}

We have found, in agreement with experiments, that the conductivity depends linearly on the density of charge carriers $n$ for $|\mu| \gg \delta \epsilon_0$, where as for $|\mu| \ll \delta \epsilon_0$, the conductivity profile has a parabolic shape. However, one may get puzzled by the fact that both (\ref{sigma-mu_eh}) and the above estimate of the minimum conductivity have been derived within the Boltzmann approach, while, as a matter of principle, the transport physics around the Dirac point is not semiclassical, but associated to relevant quantum corrections \cite{ost_etal}. A second source of difficulty is that previous semiclassical analysis, as the ones provided by the self-consistent Born approximation \cite{shon-ando} and from the numerical diagonalization of disordered hamiltonians \cite{lherbier_etal}, actually lead to minimum conductivities which are in general smaller than $4 e^2 /h$.

In order to address a solution of this confusing state of affairs, we note, as a key point, that $\delta \epsilon_0$ can be identified, within the puddle picture of disorder, with the energy scale where the semiclassical-to-quantum crossover takes place. To understand it, first recall that the energy broadening $\delta \epsilon_0$ is ultimately due to electrons delivered or captured by the substract impurities; such an energy scale is thus related to the carrier density $n_0$, which is on its turn proportional to $n_{imp}$, as already discussed at the end of Sec. III. The typical Fermi wavelength at the charge neutral point is, furthermore, $\lambda_f \sim 1/\sqrt{n_0} \sim 1/\sqrt{n_{imp}}$. In second place, we call attention to the fact that the mean distance between the scattering impurities is $\delta \sim 1/\sqrt{n_{imp}} \sim \lambda_f$, which indicates that $\delta \epsilon_0$ gives indeed a crossover energy scale for electronic transport in graphene. 

If we assume now that transport in the chemical potential range $|\delta \mu| > \delta \epsilon_0$ is to some good approximation described by the formalism developed in Sec. III, then it is clear that the analytical matching between the semiclassical (linear) and quantum (parabolic) profiles of the conductivity at the energy scale $\delta \epsilon_0$ is the essential reason for the validity of (\ref{sigma-mu_eh}). From this point of view, we also conclude that our estimate of the minimum conductivity is not quite a direct semiclassical evaluation - rather, it is the result of an extrapolation, by curve matching, of the conductivity behavior at the crossover energy scale $\delta \epsilon_0$.

\section{Conclusions}

We have introduced a statistical version of the puddle picture of 
graphene disorder, which provides a simple and straightforward phenomenological 
Boltzmann transport treatment of graphene conductivity. We have been
able to predict and verify in this way an unsuspected relationship that holds 
for the geometrical parameters that define the parabolic shape of the 
conductivity profile near its minimum value, as a function of the backgate 
voltage, and the mobility of charge carriers in the linear regime, far from the
charge neutral point.

The experimental validation of the original formulation of the puddle model \cite{hwang_etal}
is an involved issue, which goes beyond the mere fitting of measured conductivity profiles. 
Our central result, Eq. (\ref{sigma-mu_eh}), bypasses this kind of difficulty, since it does not 
rely, in principle, on specific modeling details, like the density, statistical distribution and nature 
of the disordering impurities. Fig. 2 provides, therefore, an important support for the general 
validity of the puddle picture of graphene disorder.

Natural extensions of the present work are related to the consideration 
of electron-hole asymmetry, temperature, and interaction effects. It would be
interesting to investigate if similar results can be derived for the case of clean 
suspended graphene samples, where the puddle picture of disorder is probably 
not adequate anymore.

\acknowledgments
\vspace{-0.0cm}

This work has been partially supported by CNPq. We thank Qu Fanyo,
Caio Lewenkopf and Felipe Pinheiro for enlightening discussions. 
L.M. would also like to thank the warm hospitality at ICTP, where 
part of this work was completed.


\begin{references}

\bibitem{geim_novo} A.K. Geim and K.S. Novoselov, Nat. Mater. {\bf{6}}, 183 (2007).

\bibitem{castro_etal} A.H. Castro Neto, F. Guinea,
N.M.R. Peres, K.S. Novoselov, and A.K. Geim,
Rev. Mod. Phys. {\bf{81}}, 109 (2009).

\bibitem{peres}  N. M. R. Peres, Rev. Mod. Phys. {\bf{82}}, 2673 (2010).

\bibitem{mucci-caio} E.R. Mucciolo and C.H. Lewenkopf, J. Phys.: Cond. Mat. {\bf{22}},
273201 (2010).

\bibitem{DasSarma_etal} S. Das Sarma, S. Adam, E. H. Hwang, and E. Rossi,
Rev. Mod. Phys. {\bf{83}}, 407 (2011). 

\bibitem{ryu_etal} S. Ryu, C. Mudry, A. Furusaki, and A. W. W. Ludwig,
Phys. Rev. B {\bf{75}}, 205344 (2007).

\bibitem{hwang_etal} E. H. Hwang, S. Adam, and S. Das Sarma, 
Phys. Rev. Lett. {\bf{98}}, 186806 (2007).

\bibitem{adam_etal} S. Adam, E.H. Hwang, V.M. Galistki, and S. Das Sarma,
Proc. Natl. Acad. Sci. {\bf{104}}, 18392 (2007).

\bibitem{martin_etal} J. Martin, N. Akerman, G. Ulbright, T. Lohmann,
J.H. Smet, K. Von Klitizing, and A. Yacoby, Nat. Phys. {\bf{4}}, 144 (2008).

\bibitem{zhang_etal} Y. Zhang, V.W. Brar, C. Girit, A. Zettl, and M.F. Crommie,
Nat. Phys. {\bf{5}}, 722 (2009).

\bibitem{DasSarma2_etal} S. Das Sarma, E.H. Hwang, and E. Rossi,
Phys. Rev. B {\bf{81}}, 161407 (R) (2010).

\bibitem{deshpande_etal} A. Deshapande, W. Bao, H. Zhang, Z. Zhang,
Z. Zhao, C.N. Lau, and B.J. LeRoy,
Phys. Rev. B {\bf{83}}, 155409 (2011).

\bibitem{kats_etal} M.I. Katsnelson, K.S. Novoselov, and A. K. Geim,
Nat. Phys. {\bf{2}}, 620 (2006).

\bibitem{novo_etal} K. S. Novoselov, A. K. Geim,  S. V. Morozov,  D. Jiang,
Y. Zhang, S. V. Dubonos, I. V. Grigorieva, A. A. Firsov,
Science {\bf{306}}, 666 (2004).

\bibitem{kats} M.I. Katsnelson, Eur. Phys. J. B {\bf{51}}, 157 (2006); ibid. {\bf{52}}, 151 (2006).

\bibitem{drago} D. Dragoman, Phys. Scr. {\bf{81}}, 035702 (2009).

\bibitem{trushin_etal} M. Trushin, J. Kailasvuori, J. Schliemann, A.H. MacDonald,
Phys. Rev. B {\bf{82}}, 155308 (2010).

\bibitem{cheianov_etal} A similar picture has been previously discussed in the random
network resistance model of V.V. Cheianov, V.I. Falko, B.L. Altshuler, 
and I.L. Aleiner, Phys. Rev. Lett. {\bf{99}}, 176801 (2007).

\bibitem{comment} There is, in general, some deviation of
electron-hole symmetry in graphene samples, as it is clear from
the measured values of the electron and hole mobilities.

\bibitem{chen_etal} J.-H. Chen, C. Jang, S. Adam, M.S. Fuhrer, E.D. Williams,
and M. Ishigami, Nat. Phys. {\bf{4}}, 377 (2008).

\bibitem{chen2_etal} J.-H. Chen, W.G. Cullen, C. Jang, M.S. Fuhrer, and E.D. Williams,
Phys. Rev. Lett. {\bf{102}}, 236805 (2009).

\bibitem{mccreary_etal} K.M. McCreary, K. Pi, A.G. Swartz, W. Han, W. Bao, C.N. Lau, F. Guinea, 
M.I. Katsnelson, and R.K. Kawakami, Phys. Rev. B {\bf{81}}, 115453 (2010).

\bibitem{xiao_etal} S. Xiao, J.-H. Chen, S. Adam, E.D. Williams, and M.S. Fuhrer,
Phys. Rev. B {\bf{82}}, 041406(R) (2010).

\bibitem{ost_etal} P.M. Ostrovsky, I.V. Gornyi, and A.D. Mirlin,
Phys. Rev. B {\bf{74}}, 235443 (2006).

\bibitem{shon-ando} N.H. Shon and T. Ando, J. Phys. Soc. Jap. {\bf{67}}, 2421 (1998).

\bibitem{lherbier_etal} A. Lherbier, B. Biel, Y.-M. Niquet, and S. Roche,
Phys. Rev. Lett. {\bf{100}}, 036803 (2008).

\end{references}
\end{document}